\documentclass[iop]{emulateapj}

\usepackage{epsfig}

\slugcomment{accepted, ApJ}

\shorttitle{Evidence Against Long-Lived Spiral Arms}
\shortauthors{Foyle et al.\ }

\begin{document}

\title{ Observational Evidence Against Long-Lived Spiral Arms in Galaxies}

\author{K.Foyle}
\affil{Max-Planck-Institute f\"{u}r Astronomie, K\"{o}nigstuhl 17, D-69117, Heidelberg, Germany}
\affil{Dept. of Physics \& Astronomy, McMaster University, Hamilton, Ontario, L8S 4M1, Canada}
\author{H.-W.Rix}
\affil{Max-Planck-Institute f\"{u}r Astronomie, K\"{o}nigstuhl 17, D-69117, Heidelberg, Germany}
\author{C. L. Dobbs}
\affil{Max-Planck-Institut f\"{u}r extraterrestrische Physik, Giessenbachstra\ss e, D-85748 Garching, Germany }
\affil{Universitats-Sternwarte M\"unchen, Scheinerstra\ss{}e 1, D-81679 M\"unchen, Germany}
\author{A. K. Leroy}
\affil{National Radio Astronomy Observatory, 520 Edgemont Rd., Charlottesville, VA 22903 USA; Hubble Fellow}
\author{F. Walter}
\affil{Max-Planck-Institute f\"{u}r Astronomie, K\"{o}nigstuhl 17, D-69117, Heidelberg, Germany}

\email{foylek@physics.mcmaster.ca}

\begin{abstract}
We test whether the spiral patterns apparent in many large disk galaxies should be thought of as dynamical features that are
stationary in a co-rotating frame for $\gtrsim t_{dyn}$, as implied
by the density wave approach for explaining spiral arms.
If such spiral arms have enhanced star formation (SF), observational
tracers for different stages of the SF sequence
should show a spatial ordering,
from up-stream to downstream in the corotating frame:
dense \ion{H}{1}, CO, tracing molecular hydrogen gas, 24 $\mu$m emission
tracing enshrouded SF and UV emission tracing
unobscured young stars. We argue that such a spatial ordering should be reflected
in the angular cross-correlation (CC, in polar coordinates) using all azimuthal positions among pairs
of these tracers; the peak of the CC should be offset from zero, in different directions inside and outside the corotation radius.
Recent spiral SF simulations by Dobbs \& Pringle,
show explicitly that for the case of
a stationary spiral arm potential such angular offsets between gas and young
stars of differing ages should be observable as cross-correlation offsets.
We calculate the angular cross-correlations for different
observational SF sequence tracers in 12 nearby
spiral galaxies, drawing on a data set with high quality maps of
the neutral gas (\ion{H}{1}, THINGS), molecular gas (CO, HERACLES) along with 24 $\mu$m emission ({\it Spitzer}, SINGS); we include FUV images ({\it GALEX}) and 3.6 $\mu$m emission ({\it Spitzer}, IRAC) for some galaxies, tracing aging stars and longer timescales.
In none of the resulting tracer cross-correlations for this sample
do we find systematic angular offsets, which would be expected for a stationary
dynamical spiral pattern of well-defined pattern speed. This result indicates that spiral density waves in their simplest form are not
an important aspect of explaining spirals in large disk galaxies. 
 \end{abstract}

\keywords{galaxies: general ---}
\section{Introduction}
 The nature of a coherent spiral structure that extends over a large portion of the galaxy is still not fully understood.  While there are many variations in theories to explain such a structure,  typically self-excited models for spiral structures can be divided into two groups based on the longevity of the spiral pattern.   A long-lived quasi-stationary spiral structure theory has been developed largely using analytical studies. The 
spiral features are attributed to quasi-steady global modes of the disk (Lin \& Shu 1964, 1966; Lynden-Bell \& Kalnajs 1972; Bertin et al.\ 1989a, 1989b, etc.). The other approach, which has largely been based on simulations, purports that spirals are short-lived, recurrent, transient patterns that originate from recurrent gravitational instabilities (Goldreich \& Lynden-Bell 1965; Julian \& Toomre 1966; Toomre 1981; Sellwood \& Carlberg 1984; Sellwood 2010).   It remains a challenge to decipher which of these descriptions best describes spiral structure.     The interested reader can refer to excellent reviews in Binney \& Tremaine (2008) and Sellwood (2010) for more information.   In these reviews it is explained that it has proven very difficult to decipher which of these models best describes observed galaxies and it 
may be that a combination of these models is required. It is clear that careful comparisons between observations and 
simulations are needed in order to disentangle these theories. 

In this work, we examine simulations and observations for evidence of a long-lived, quasi-stationary spiral structure based on the predictions of such a model.   As originally explained by Roberts (1969, hereafter R69), if the spiral pattern is a quasi-stationary, dynamical pattern in a corotating frame and SF is occurring at an enhanced rate in the arms, then there should be a temporal sequence of events, as material streams in and out of this spiral pattern.  

As gas passes through the minimum of the spiral potential, it gets compressed and molecular clouds form.  This is  followed by SF, leading initially to dust-enshrouded young stars and then to unobscured young stars and clusters. If this SF sequence of events were to occur in a steady state, then it would translate into a set of spatial offsets for tracers of different stages of this SF sequence, with the size of the offsets reflecting both the time difference between these stages and the relative velocity between the material and the spiral pattern.  In this way, observed spiral arms are produced by the continual triggering of SF through compression in the gas peak (Lubow et al. 1986). Even if the spiral structure is not directly shock triggering SF (e.g., Foyle et al. 2010), provided there is an appreciable amount of SF in the arms, one still expects bright stars to be observed downstream from the arms and this has been noted in several cases (e.g., Tamburro et al. 2008; Egusa et al. 2009).

\begin{figure}

\centering

 \includegraphics[trim=10mm 0mm 10mm 0mm, clip,width=80mm]{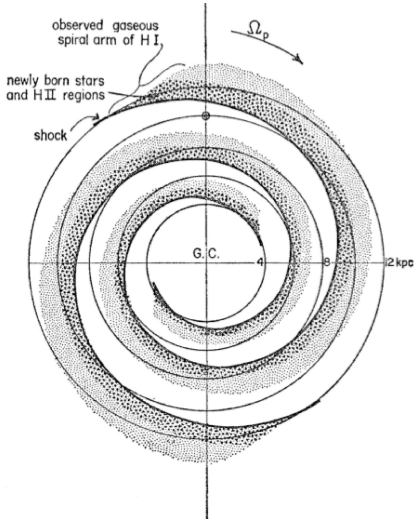}

\caption[Illustration of the ordering of gas and stars in a long-lived spiral structure and results of a simulation showing the ordering of stellar cluster ages.]{Illustration of the relative location of gas shock, sharp \ion{H}{1} peak and newly formed stars in \ion{H}{2} regions in the gaseous spiral arms of a trailing spiral pattern (top; Figure 7 of R69).}

\label{r69}

\end{figure}

Roberts et al.\ (1975) found observational evidence for such offsets from a sample of 24 galaxies and posited that the shock strength and spiral strength could account for the ordering of spiral disks into Hubble type and luminosity class.  Figure~\ref{r69} taken from R69 illustrates the  positions of the gas shock and young stars inside corotation of a trailing spiral pattern and illustrates the predicted offsets.  Using a rigidly rotating spiral potential in hydrodynamical simulations, Gittins \& Clarke (2004) showed that one can find three spiral patterns each tracing a different stage in this process.  The first is traced by the stellar mass density and the location of the potential minima.  The second spiral pattern is traced by the dust, which is known to be in locations of compressed gas and the third marks the location of the young stars.  Due to the finite time for SF, the young stars should be downstream of the dense gas in this picture. The location of the compressed gas relative to the minimum is more complex; traditionally it is located upstream of the minimum, but the location of the shock depends on the sound speed of the gas and can be downstream of the minimum in a cold, or multiphase medium (Wada 2008; Dobbs 2007).
 
\begin{figure}

\centering

 \includegraphics[trim=5mm 10mm 10mm 10mm,clip,width=40mm]{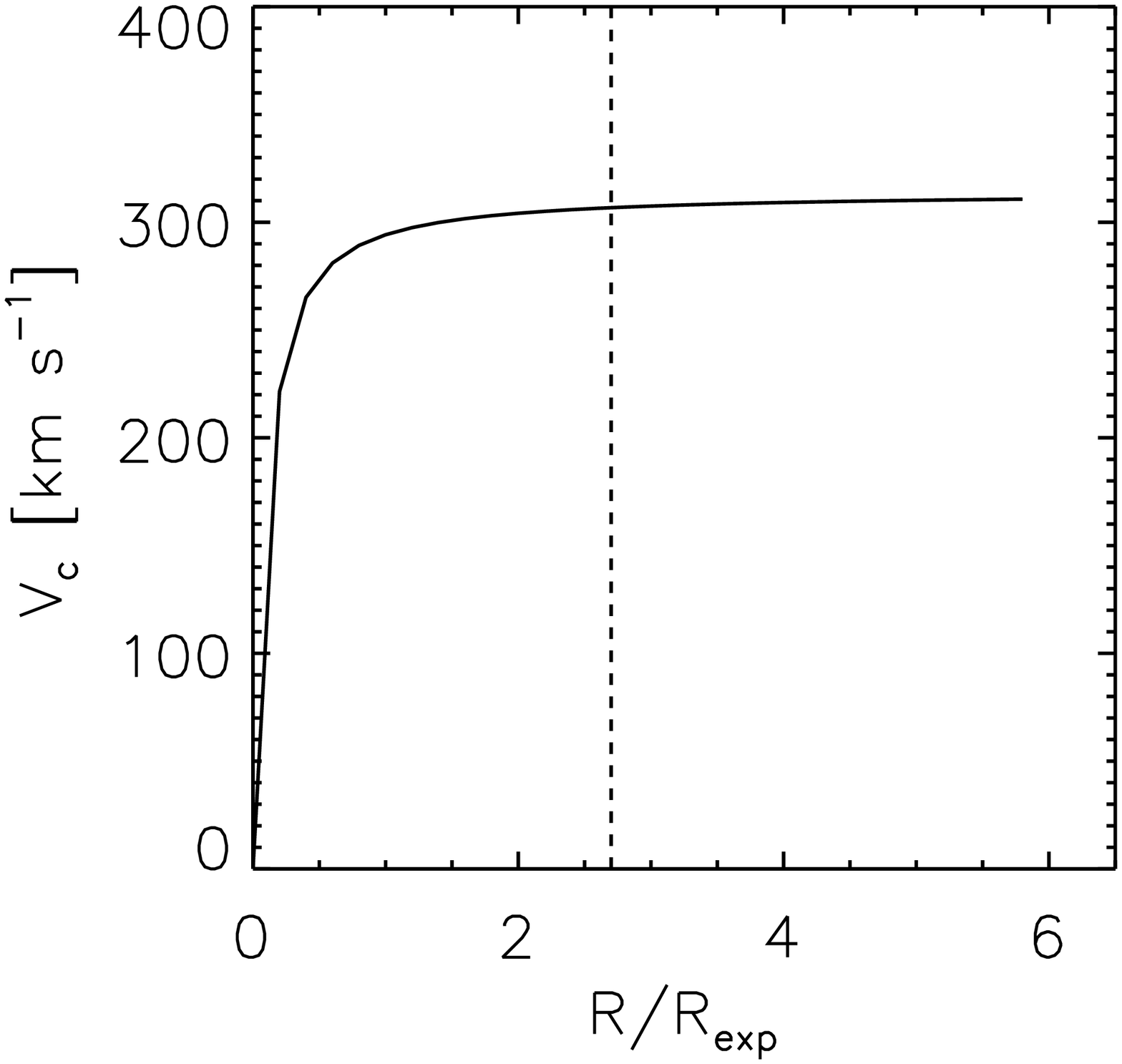}
 \includegraphics[trim=5mm 10mm 10mm 10mm,clip,width=40mm]{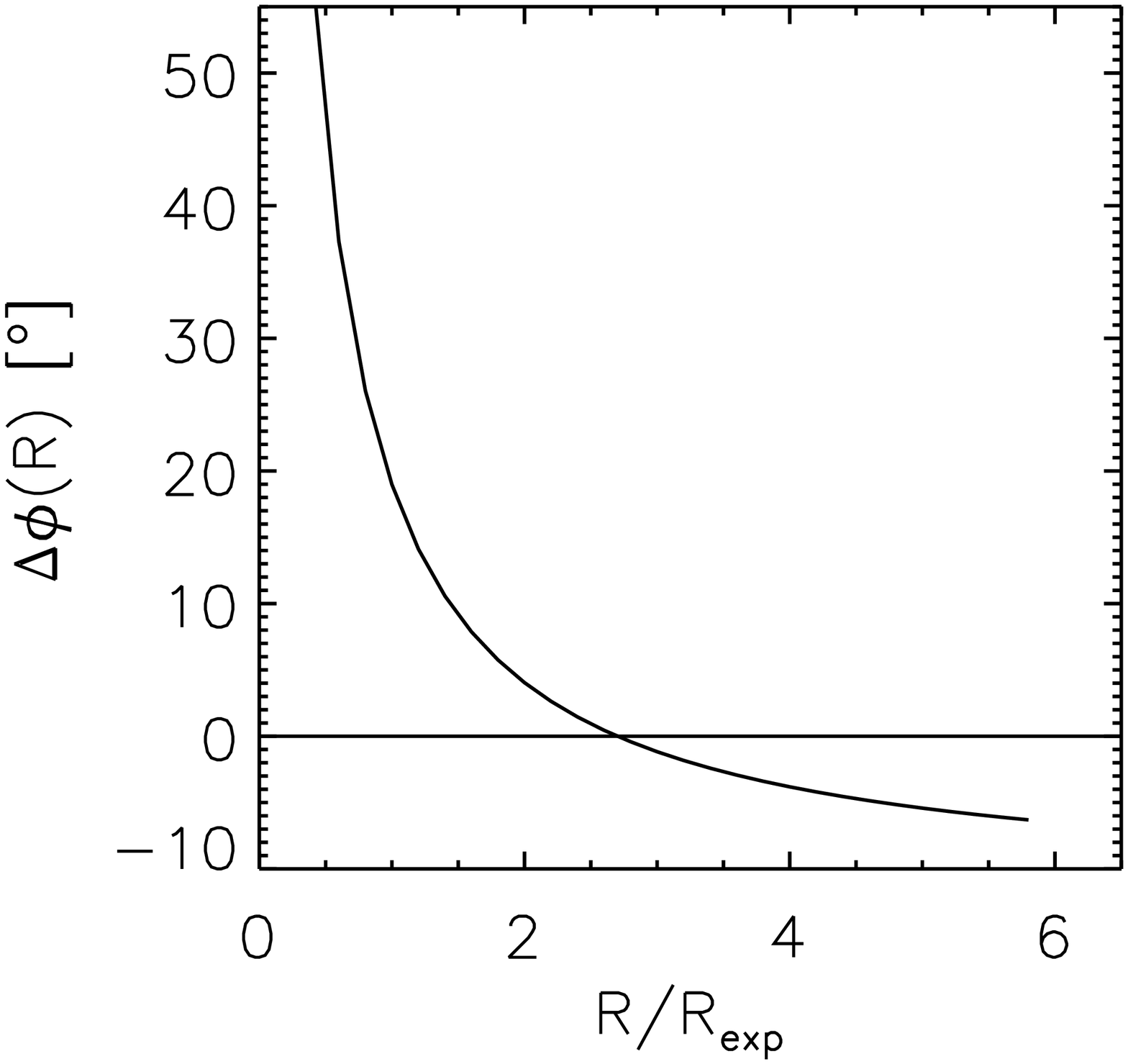}
   \includegraphics[trim=5mm 10mm 10mm 10mm,clip,width=40mm]{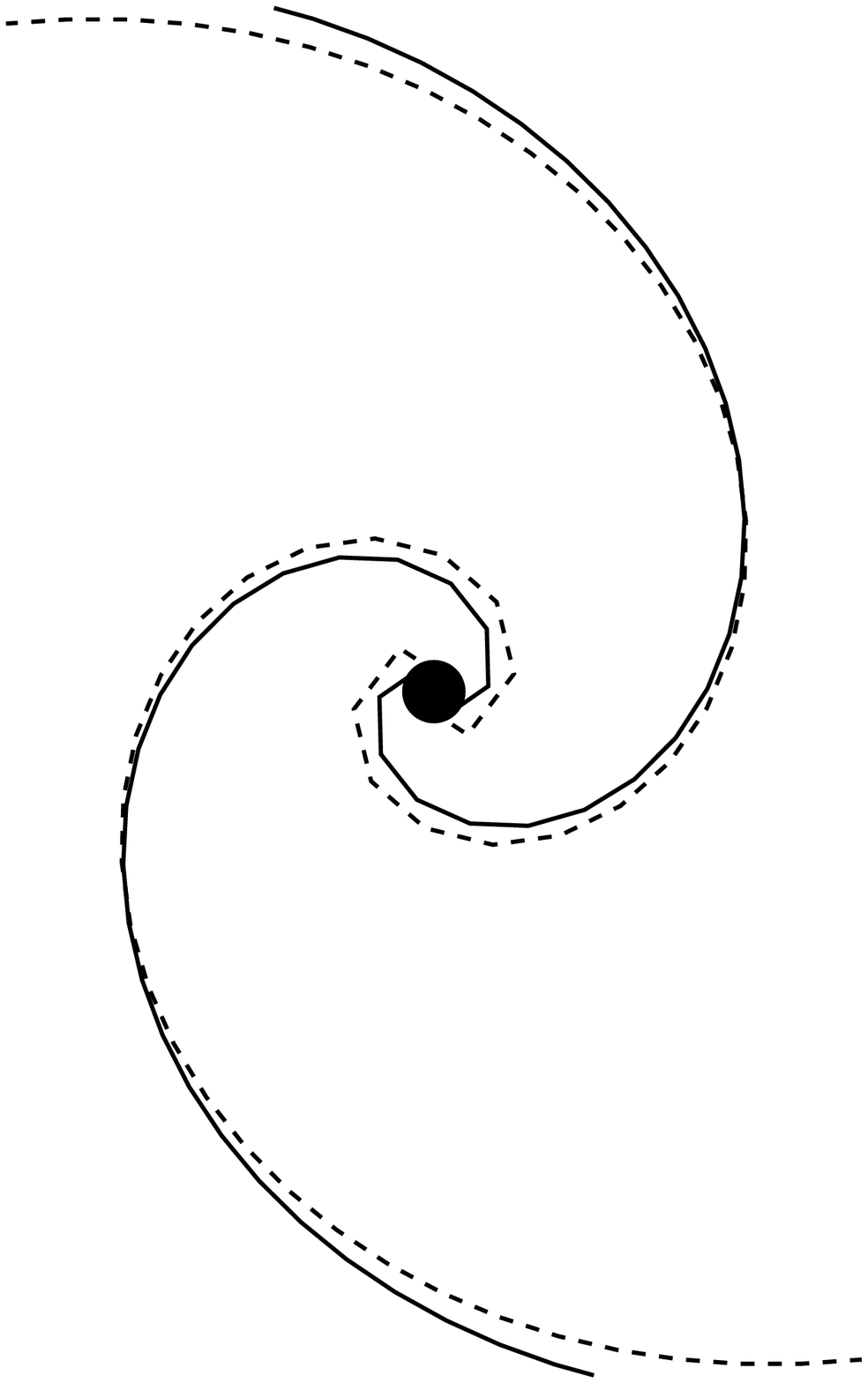}

\caption[Toy model properties showing the rotation curve, azimuthal offsets and relative positions of SF tracers in the spiral arms.]{Toy model illustration showing the relative position between the gas and a recent SF tracer  with an onset time of 3 Myr between the two.  The rotation curve of the galaxy is modeled with an $\arctan$ function (upper left) and the corotation of the spiral pattern is set at 2.7R$_{\rm exp}$ (dashed line).  With a t$_{SF}$ of 3 Myr, Eq.~\ref{mt} determines the radial profile of the angular offsets between the two tracers (upper right).  The relative positions of the spiral patterns of each tracer are shown below.}

\label{toyoff}

\end{figure}

 Figure~\ref{toyoff} shows what would be expected using a simple toy model of a two-arm spiral pattern with a corotation radius of 2.7R$_{\rm exp}$ ({\it i.e.} Kranz et al.\ 2003).  We consider a gas and recent SF tracer with an onset time of 3 Myr between the two.  The upper right panel of Figure~\ref{toyoff} shows the radial profile of the angular offsets between them calculated from Eq.~\ref{mt}, discussed in the following section, using the rotation curve shown on the left.  In the inner regions large positive offsets are expected and beyond corotation the offsets become negative.  The lower panel shows the position of the tracers in the spiral arms.  The variation in the angular offsets of the tracers with radius, produces a spiral pattern that is effectively more tightly wound. The offsets are small enough that the arm patterns of the two tracers will likely overlap, depending on the amount of dispersion and width of the tracers.  This could make by-eye determinations of offsets challenging.

A number of observational studies have looked for qualitative evidence for
angular offsets of SF tracers.  Early observations showed dust
lanes on the inside part of the arms with respect to \ion{H}{2} regions ({\it e.g.}, Lynds
1970).   Mathewson et al.\ (1972) and Rots (1975) found that atomic hydrogen was offset from optical arms and dust lanes in M51 and M81 respectively.   A detailed study of M33 by Humphreys \& Sandage (1980), however, presented a more complex picture.   While two of the arms did show dust lanes on the inner edge of the arm and bright young stars on the outer part of the arm, the third arm showed no such features or ordering.

Ideally one would like to examine offsets between the gas and young stars.  Due to the lack of atomic hydrogen (\ion{H}{1}) in the inner part of the disk and the fact that the molecular gas forms giant molecular clouds out of which stars form, it is natural to look for offsets between CO and H$\alpha$ or 24 $\mu$m emission (with both H$\alpha$ and 24 $\mu$m emission being tracers of recent SF, {\it e.g.}, Calzetti et al.\ 2007).  A number of studies have looked for such offsets and examined the streaming motions near the arms, especially in M51 and M81 ({\it e.g.}, Vogel et al.\ 1988; Rand \& Kulkarni 1990; Lord \& Young 1990; Garcia-Burillo et al.\ 1993; Rand 1995; Loinard et al.\ 1996; Shetty et al.\ 2007; Egusa et al.\ 2009).  However, all such studies used by-eye determinations of these offsets by selecting individual patches of a single arm.  Such methods can easily introduce potential biases.

Tamburro et al.\ (2008, hereafter T08) developed an algorithmic technique to measure offsets between SF tracers by locating the peak of the cross-correlation function between \ion{H}{1} and 24 $\mu$m emission.  This also allowed for a measure of the timescale of going from \ion{H}{1} atomic gas to dust-enshrouded massive stars (as traced by 24 $\mu$m emission).    Their findings were in agreement with the simple prescription of R69.  They found corotation radii at $\sim$2.7r$_{\rm exp}$, which is consistent with other works ({\it e.g.}, Kranz et al.\ 2003) and short timescales of 1-4 Myr for the timescales of SF emerging from \ion{H}{1}.  Indeed the angular offsets were found to be very small (5$^{\circ}$) between the tracers.  T08's work is the first to approach offset measurements using an algorithmic method as opposed to by-eye determinations and they used the highest resolution images to date.  Our method is based on the one developed by T08 and we describe it in greater detail in \S2.

Recently, Dobbs \& Pringle (2010, hereafter DP10) have looked in detail at the distribution of cluster ages across spiral arms in realistic simulations.  In their sample, they included a galaxy with a fixed spiral potential which mimics a long-lived spiral with a constant pattern speed.  The top left panel of Figure~\ref{DB} shows the cluster age distribution in this galaxy.  If one compares this to the illustration of R69 (Figure~\ref{r69}), one sees that the distribution is more complex in these simulations.  However, the clusters are still ordered in the way expected.  Previous studies by Dobbs et al.\ (2006) have also shown that the distribution of the giant molecular clouds has a sharp edge on the upstream side of the arms where fresh material is flowing in and a dispersed, smooth distribution on the downstream side.

However, imposed stationary potentials are not realistic. DP10 also included three spiral structures without external potentials, in which the spiral structure forms naturally. They included a flocculent spiral as well as a barred and interacting galaxy. Unlike imposed stationary potentials, the active potentials do not produce a spatial separation between the gas and SF tracers. Dobbs \& Bonnell (2008) have shown that the stars and gas are coincident with the spiral potential minima, because gas accumulates in the potential minima as the spiral structure forms. The spiral arms in the flocculent model of DP10 evolve over timescales of 100 Myr, which is typical for spirals with recurrent spiral structure (Sellwood \& Carlberg 1984; Fujii et al. 2011; Sellwood 2010). 
The spiral patterns found with active potentials have no fixed pattern speed or corotation radius. 
Longer lived transient spirals, where spiral arms evolve over a few hundreds of Myr are possible ({\it e.g.} Thomasson et al. 1990), and may also lead to offsets similar to the static potential case. However such structures are not readily produced, at least in numerical simulations (Sellwood 2010), and we did not include this scenario in our simulated galaxies.


In this work, we look for evidence for a systematic ordering of cold gas and young stars, which would be predicted by a long-lived spiral pattern with a constant pattern speed.  We first examine the four simulations of DP10 for angular offsets between the gas and stellar clusters and then examine a sample of observed galaxies for offsets between various gas and SF tracers. In \S2 we describe the algorithmic method developed by T08 to measure angular offsets between gas and SF tracers.  In \S3 we use this method to measure angular offsets in the simulations of DP10, which cover a range of spiral structure formation mechanisms including a stationary spiral potential, bar potential, interacting galaxy and transient spiral structure.  In \S4 we turn to observations and present our sample of 12 nearby spiral galaxies with high-quality maps of cold gas and SF tracers including \ion{H}{1} (THINGS), CO (HERACLES), 24 $\mu$m (SINGS), UV ({\it GALEX}) and 3.6 $\mu$m (IRAC).  Using the same method employed for the simulations, we measure angular offsets between these observed tracers.  In \S5 we compare the observed cross-correlation functions and angular offsets to those in the simulations and assess the predictions of the model of a long-lived spiral structure with a constant pattern speed. In \S6 we present our conclusions.

\section{Measuring Angular Offsets}
We seek to algorithmically measure angular offsets between the gas and SF tracers in both simulations and observations, which are predicted by a long-lived, quasi-stationary, spiral structure theory.  We use the method of Tamburro et al.\ 2008, which was developed to measure offsets between \ion{H}{1} and 24 $\mu$m emission.  Angular offsets between \ion{H}{1} and 24 $\mu$m emission allow one to measure the timescale between gas compression and massive SF because peaks of \ion{H}{1} column density are thought to be the sites of molecular cloud formation and the 24 $\mu$m emission traces the young dust enshrouded stars. We use this method to measure angular offsets between the gas and young stellar clusters in the simulations of DP10 and also extend T08's study to measure angular offsets between CO, which traces the molecular gas with 24 $\mu$m emission.   This provides a more direct measure than that of \ion{H}{1} and 24 $\mu$m emission. While T08 also used BIMA SONG CO maps (Helfer et al.\ 2003) to measure angular offsets between the molecular gas and 24 $\mu$m, the low sensitivity of these images provided too few points to make accurate estimates of the timescale.   We also include the cross-correlation of the gas tracers with UV images and 3.6 $\mu$m images which probe longer timescales.  We first describe the method of T08 in detail.

\subsection{Method of Tamburro et al.\ 2008}
T08 measured angular offsets by locating the peak of the cross-correlation function between \ion{H}{1} and 24 $\mu$m emission in radial annuli for 14 disk galaxies. 

  The time between these two events, $t_{HI \rightarrow 24\mu m}$, the local circular velocity, $v_{c}(r)$, and a pattern speed, $\Omega_{p}$, for the spiral arms, lead to an angular offset between these two tracers of:

\begin{equation}
\label{ft}
\Delta \phi(r)=(\Omega(r)-\Omega_{p})t_{HI \rightarrow 24\mu m}.
\end{equation}

Thus, this method is independent of the number of arms and shape of the spiral pattern, but it does assume a fixed pattern speed.  The angular offsets should vary with radius.  When the disk rotates faster than the pattern (inside corotation) we expect $\Delta \phi >$ 0 and when the pattern rotates faster than the disk (beyond corotation) we expect $\Delta \phi <$ 0.  At corotation $\Omega(R_{cor})=\Omega_{p}$, the sign changes and $\Delta \phi =$ 0.

In order to measure $\Delta \phi$, T08 divided the images into radial annuli and cross-correlated them to find the relative offset with the best match.  The best match between the two images is found by minimizing the following quantity as a function of the shift, $l$ (lag):
\begin{equation}
\chi^{2}_{x,y}(l)=\Sigma_{k}[x_{k}-y_{k-l}]^{2}.
\end{equation}
The sum is carried out over all N elements of $x$, $y$, where $x$ and $y$ refer to the two tracers at a given radius as a function of azimuth, $\phi$, such that:
\begin{equation}
x_{k}=f_{HI}(\phi_{k}|\hat{r}) \text{ and } y_{k-l}=f_{24\mu m}(\phi_{k-l}|\hat{r}).
\end{equation}
By minimizing $\chi^{2}$ one maximizes the following quantity:
\begin{equation}
cc_{x,y}(l)=\Sigma_{k}[x_{k} \times y_{k-l}],
\end{equation}
defined as the cross-correlation coefficient.  T08 used the normalized version such that:
\begin{equation}
\label{ccl}
cc_{x,y}(l)=\frac{\Sigma_{k}[(x_{k}-\bar{x})(y_{k-l}-\bar{y})]}{\sqrt{\Sigma_{k}(x_{k}-\bar{x})^{2}\Sigma_{k}(y_{k}-\bar{y})^{2}}},
\end{equation} 
where $\bar{x}$ and $\bar{y}$ are the means of $x$ and $y$ respectively.  In this way, perfectly identical patterns have a cross-correlation coefficient of unity and highly dissimilar patterns have a value much less than one.  At each annulus, T08 searched for a local maximum of the $cc(l)$ around $l \simeq 0$.  The location of the maximum, $l_{max}$, defines the angular offset between the two tracers such that $\Delta \phi(r) =-l_{max}(r)$.  If the value of the $cc(l)$ peak was less than 0.3, the point was rejected.

With the angular offset, $\Delta \phi(r)$, and the rotation curve, $v_{c}(r)$, one can write Eq.~\ref{ft} as: 
\begin{equation}
\label{mt}
\Delta \phi(r)= \Biggl( \frac{v_{c}(r)}{r}-\Omega_{p} \Biggr) t_{HI \rightarrow 24\mu m}.
\end{equation}
T08 used $\chi^{2}$ fitting of the above with the measured angular offsets and derived best fits for $t_{HI \rightarrow 24\mu m}$ and $\Omega_{p}$.  This method can be used for any two tracers.  We first employ this method on the simulations of DP10 and examine which models fit the predictions of R69.   We then use this method to repeat the work of T08 as well as look for offsets between 24 $\mu$m and the higher sensitivity CO imaging from HERACLES, which traces the molecular gas.  We also examine UV and 3.6 $\mu$m for offsets with respect to the molecular gas.

\section{Angular Offsets in Simulations}
\begin{figure}

\centering

 \includegraphics[trim=0mm 0mm 0mm 0mm, clip,width=80mm]{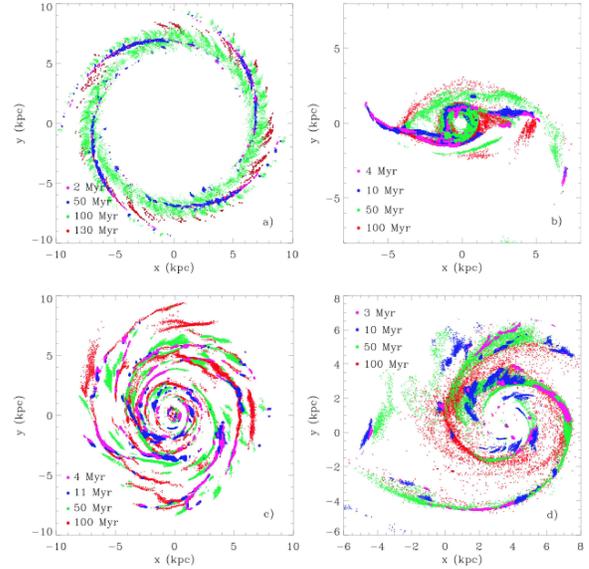}

\caption[Distribution of stellar cluster ages in the spiral arms of four simulations.]{Distribution of stellar cluster ages (shown in different colors) in the spiral arms of a long-lived spiral potential with a constant pattern speed (top left), a barred galaxy (top right), a self-excited flocculent galaxy (bottom left), and a galaxy interacting with a companion (bottom right; Figure 2 of DP10).  While the ordering of the stellar clusters is systematic in the long-lived spiral, the other cases show a much more complex distribution.}

\label{DB}

\end{figure}

DP10 have recently simulated four different models of spiral structure and compared the distribution of stellar clusters of different ages.  Their simulations included a stationary spiral potential which mimics a long-lived spiral structure with a constant pattern speed, a barred galaxy, an interacting galaxy like NGC 5194 and a transient spiral produced using the spiral potential from the models of Sellwood \& Carlberg (1984).  As we saw in \S1,  the long-lived spiral structure was simulated using a stationary potential and indeed the ordering of the stellar cluster ages fit the predictions of R69 and others (see Figure~\ref{r69}).   Since the long-lived structure requires a stationary potential in order to ensure its longevity, the other three cases may be more realistic (see Figure~\ref{DB} for all four cases).  In these cases the ordering is much more complex.  For the flocculent galaxy, each different segment of the spiral arms contains clusters of different ages.  The interacting galaxy has an almost incoherent distribution. 

We have used the same simulations presented in DP10 and analyzed them for angular offsets between the gas and stellar clusters of different ages in the way designed by T08 for observations.  For different snapshots we convert the gas and stellar cluster onto an R-$\phi$ grid and cross correlate the annuli and locate the peak of the cross-correlation function as described in \S2.  We show here the results of the cross-correlation between the gas and 100 Myr clusters in all four simulations.  The gas and the 100 Myr clusters represent timescales close to those observed between H$_{2}$ and the UV, which we will examine in \S5.


\subsection{Stationary Spiral Potential}
\begin{figure}

\centering

 \includegraphics[trim=20mm 0mm 0mm 100mm, clip,width=120mm]{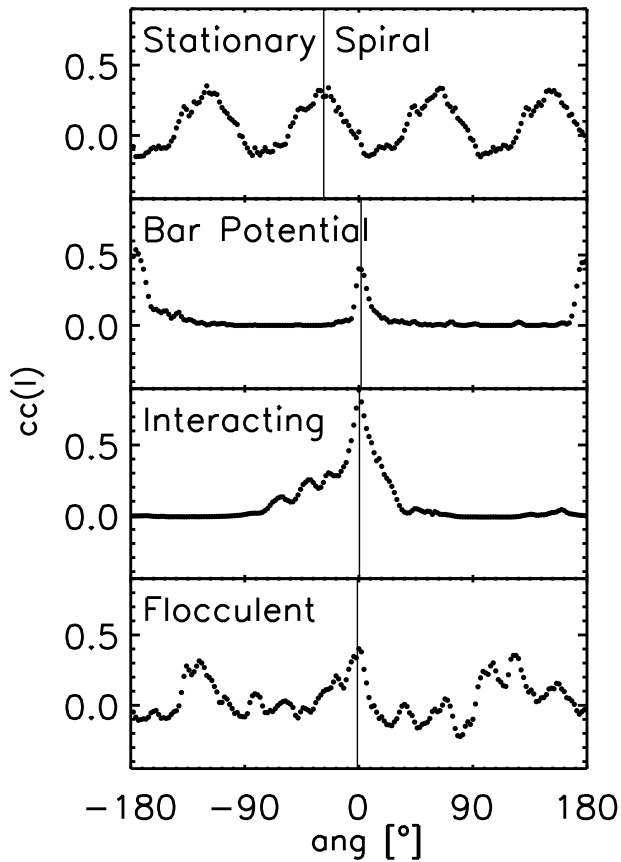}

\caption{Cross correlation function between the gas and the 100 Myr clusters for the simulated stationary spiral potential (top panel), bar potential (second from top), interacting spiral (third from top) and flocculent spiral (bottom) at a fiducial radius of 7 kpc in each case.  The cross-correlation function of the stationary potential shows four equally spaced peaks due to the self-similarity of the four-armed spiral structure.  In the other cases, the central peaks are less well-defined and show very small, if any, offsets.}

\label{DBcc}

\end{figure}

\begin{figure}

\centering

 \includegraphics[trim=0mm 0mm 0mm 100mm, clip,width=80mm]{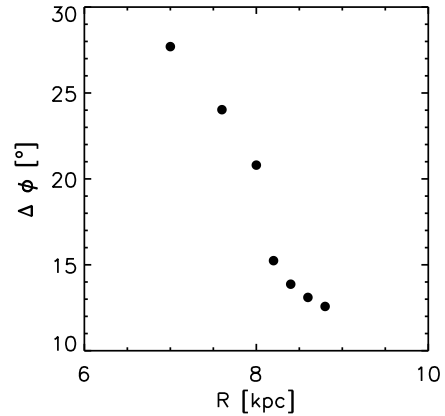}

\caption{Radial profile of the angular offsets based on the peak of the cross-correlation function between the gas and 100 Myr cluster particles for the stationary spiral potential.  The offsets show a systematic trend going from high to low values.  The gas and star particles in this simulation do not extend beyond the corotation of the spiral potential.  Thus, we are unable to currently test whether the offsets cross zero at this point.  }

\label{radang}

\end{figure}

The stationary spiral potential is an imposed potential with a constant pattern speed.  Thus, we expect an ordering of the clusters and gas to be in the way predicted.  The pattern has a corotation radius at 11 kpc.  The  upper panel Figure~\ref{DBcc} shows the cross-correlation function between the gas and the expected distribution of the 100 Myr stellar clusters at a fiducial radius of 7 kpc.  The galaxy has four arms, so there are four peaks in the cross-correlation function due to the self-similarity of the pattern.  At each radius, we fit the central peak with a four degree polynomial, as was done by T08.  The peak of the polynomial locates the angular offset.  Figure~\ref{radang} shows the chosen angular offset between the gas and 100 Myr clusters at a series of radial annuli.  As one moves out in radius the offsets decrease in value as we saw with the toy model in \S1.  Unfortunately, the stellar clusters do not extend beyond corotation so we are unable to confirm that the sign of the offsets changes beyond corotation.

\begin{figure}

\centering

 \includegraphics[trim=50mm 30mm 0mm 200mm, clip,width=100mm]{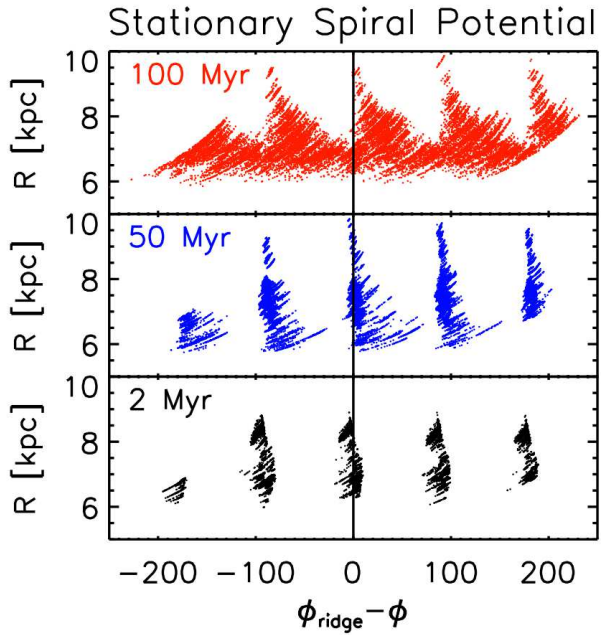}

\caption{Angular distance of clusters of varying ages to the gaseous spiral arms in the stationary spiral potential simulation.  By fitting a ridge line to a spiral arm in the gas ($\phi_{ridge}$), we measure the angular distance to each age of stellar cluster (2 Myr: black; 50 Myr: blue; 100 Myr: red).  As the clusters age they drift further downstream with respect to the arm, as expected in the stationary, spiral potential model.}

\label{DBridge}

\end{figure}

Since the arms are well-defined in the stationary potential we fit straight lines to the gas arms in logR-$\phi$ and defined a ridge line to the arm.  At each radial position the arm is at a position $\phi_{ridge}$.  We then measure the distance of the clusters with respect to the arm.  Figure~\ref{DBridge} shows stellar clusters at 2, 50, and 100 Myr and their angular distance to the ridge line.  We have only fit one arm so the self-repeating pattern is due to the other arms.  One notes that the older clusters have drifted further downstream from the arms ({\it i.e.} ($\phi_{ridge} -\phi$) increases for clusters of older ages).  Both these findings and the systematic ordering of  the angular offsets fit the predictions of a long-lived, stable, spiral structure.  This is not surprising because in these simulations such a structure is imposed.  This exercise shows that if such a structure exists in nature, we should be able to detect offsets between gas and SF tracers using our analysis.

We now turn to the other simulations, where the spiral structure has formed naturally in the simulations and is not fixed with a stationary potential.  In these cases, however, we do not expect a constant pattern speed for the spiral pattern, which is one of the assumptions in the R69 model.

\subsection{Barred Galaxy}

The second panel from the top of Figure~\ref{DBcc} shows the cross-correlation functions of the gas and the expected distribution of the 100 Myr stellar clusters for the barred potential at a fiducial radius of 7 kpc.  Here we find two cross-correlation peaks due to the two arms in this galaxy, but the cross-correlation peaks are small and, in many cases, are below the 0.3 cutoff value used by T08 and which we also adopt for the observed galaxies.   Despite these low $cc(l)$ values, the peak is well-defined so we are able to determine the offset angles.  We find that the offsets are very small and, at all radii, are less than a degree. One concern is that by 100 Myr the ordering of the particles has been disrupted.  In comparison to the other simulated galaxies, the bar primarily extends over smaller radii, and the orbital times are shorter. By 100 Myr, the particles have made at least one and possibly two passages around the galaxy, disrupting the ordering of the gas and star particles.  Due to this, we also examined star particles at early times, before complete passages.  For the 50 Myr clusters the offsets were easier to detect and larger, but still quite small with the largest being $\sim$ 5$^{\circ}$.  A further challenge in the barred galaxy is the highly elliptical orbits of the particles, which make the transitions more difficult to detect using the cross-correlation method, which relies on the azimuthal angle of the particles. In general, we find that offsets can be detected for the barred galaxy, but that they are quite small and the ordering of the particles can be easily disrupted.

\subsection{Interacting Galaxy}

The second panel from the bottom of Figure~\ref{DBcc} shows the cross-correlation functions for the gas and the expected distribution of the 100 Myr stellar clusters of the interacting galaxy, which is very similar to NGC 5194, at a fiducial radius of 7 kpc.  In this case, we find only a single, broad, uneven cross-correlation peak.  The measured angular offsets show no trend and in most cases the angular offsets are close to zero and thus we do not show them graphically.    Since this simulated galaxy is very similar to NGC 5194 (see DP10 for further details), it is interesting to compare the cross-correlation function with the observations, which we will do in the following section.  

\subsection{Transient, Flocculent Spiral}

Finally we also examine the cross-correlation function of the transient, more flocculent spiral structure that used the spiral potential from the models of Sellwood \& Carlberg (1984).  The  bottom panel  of Figure~\ref{DBcc} shows the cross-correlation function between the gas and the expected distribution of the 100 Myr stellar clusters at a fiducial radius of 7 kpc.  At this radius and all other ones no obvious cross-correlation peak is found, suggesting no preferred position between the gas and clusters.  We were unable to properly fit a peak and measure angular offsets and thus, we do not show the radial variation here.  

\subsection{Summary of Numerical Models}
If a long-lived spiral structure with a constant pattern speed accurately describes spiral structure, than we should be able to detect angular offsets between gas and SF tracers as we have seen with the imposed stationary potential of DP10.  However, if a stationary potential is not imposed, simulations of spiral structure do not show a systematic ordering of angular offsets between the gas and stellar clusters in fitting with the model predictions of a long-lived, stable, spiral structure.   In the case of the barred galaxy, the cross-correlation peak could be fit, but the values were very low and the angular offsets measured, were small. In the other two cases, the offsets showed no radial variation and were close to zero.  We now turn to our sample of observed galaxies and compare our findings.

\section{Angular Offsets in Observed Galaxies}
We chose 12 galaxies in common with the sample of T08 that have coverage in THINGS (Walter et al.\ 2008) and SINGS (Kennicutt et al.\ 2003).  Eight of these galaxies also have coverage in HERACLES (Leroy et al.\ 2009).   For three galaxies we also included FUV images from {\it GALEX} (Gil de Paz et al.\ 2007) and 3.6 $\mu$m images from IRAC (Kennicutt et al.\ 2003).  The sample presents a mixture of both grand design and more flocculent spirals, which are similar to the four simulated galaxies of DP10.

\begin{figure}

\centering

 \includegraphics[trim=20mm 10mm 35mm 100mm, clip,width=110mm]{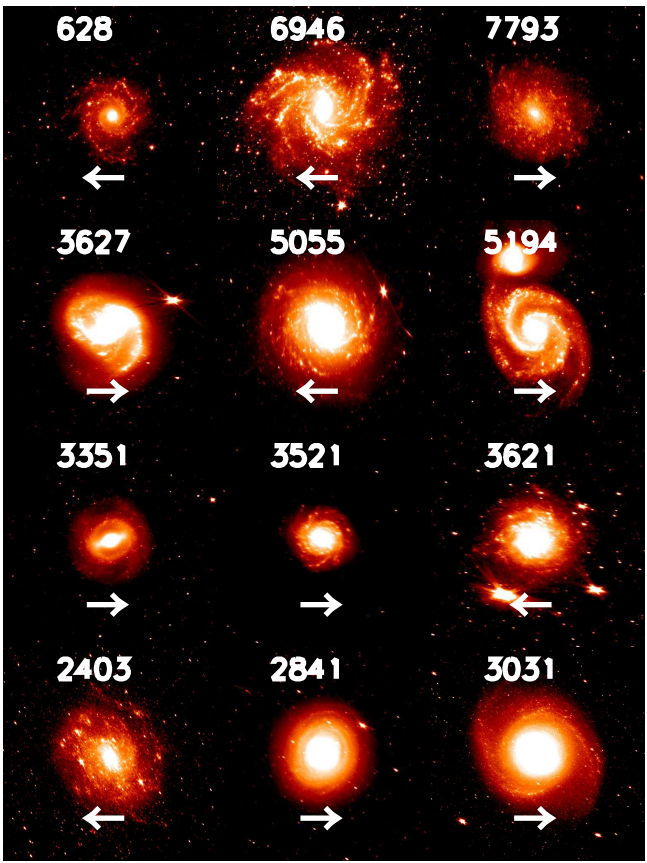}

\caption[Sample of 12 galaxies.]{Sample of 12 galaxies (numbers correspond to NGC numbers).  We show the deprojected 3.6 $\mu$m images with arrows denoting the direction of rotation based on the assumption of a trailing spiral pattern.}

\label{gals}

\end{figure}

The galaxies were deprojected according to the values of T08, with the exception of NGC 2841, NGC 3521 and NGC 5194 (see Table 1).  For NGC 5194, we used the values adopted by Leroy et al.\ (2008).  NGC 2841 and NGC 3521 were deprojected using GALFIT (Peng et al.\ 2002) and are roughly equivalent to those found in HyperLeda Extragalactic Database.   Figure~\ref{gals} shows deprojected 3.6 $\mu$m images for our sample as well as the direction of rotation, which is defined by assuming a trailing spiral pattern.

The images were aligned to the THINGS astrometric grid and were degraded to a common resolution of either 6$''$ for our analysis of \ion{H}{1} and 24 $\mu$m images or 13$''$ for our analysis of CO, 24 $\mu$m, UV and 3.6 $\mu$m images.  Foyle et al.\ (2010) describes in detail the processing of the images.  

\begin{table}

\begin{center}

\caption{Properties of 12 Sample Galaxies}

\begin{tabular}{ccccccc}

\hline\hline

Name & Inclination &  P.A. & $D$ & $r_{\rm exp}$ & V$_{max}$ \\

& [$^{\circ}$] & [$^{\circ}$] & [Mpc] &$[']$ & [km s$^{-1}$]\\
NGC 628 & 7 & 20 & 7.3 &1.1 &220 \\

NGC 2403 & 63 & 124 & 3.22 & 1.3 & 128  \\

NGC 2841 & 63 &148 &14.1 & 0.92 & 331 \\

NGC 3031 & 59 & 330  & 3.63  & 3.62 & 256\\

NGC 3351 & 41 & 192 & 9.33 & 0.86 & 210 \\

NGC 3521 & 65 & 162 & 10.05 & 0.74 & 242\\

NGC 3621 & 65 &345 &6.64 & 0.8 & 144\\

NGC 3627 & 62 & 173 & 9.25  & 0.95 & 204\\

NGC 5055 & 59 & 102 & 7.82 & 1.16 & 209  \\

NGC 5194 & 20 & 172 & 7.77 & 1.39 &242  \\

NGC 6946 & 33 & 242 & 5.5 & 1.73 & 201 \\

NGC 7793 & 50 & 290 & 3.82 & 1.16 & 109 \\

\hline

\end{tabular}

\tablecomments{Sample properties: the inclination and position angles used to deproject the galaxies, their adopted distance, scale length and maximum amplitude of their rotation velocity.  With the exception of NGC 2841, NGC 3521 and NGC 5194, all values are those from T08.}
\end{center}

\end{table}


\begin{figure*}

\centering

 \includegraphics[trim=35mm 35mm 10mm 150mm, clip,width=140mm]{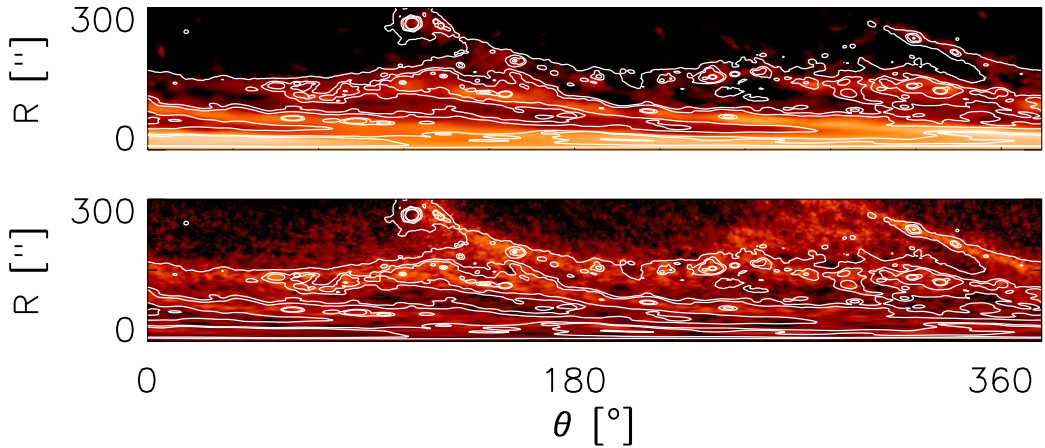}

\caption[\ion{H}{1} and CO Emission of NGC 5194 in polar coordinates.]{\ion{H}{1} (bottom) and CO (top) emission of NGC 5194 in polar coordinates with contours of the 24 $\mu$m emission overlaid.  It is clear from these overlays that any angular offsets between both sets of tracers must be small.}

\label{m51}

\end{figure*}

 After initial processing, the images are translated from rectangular to polar coordinates, $(r,\phi)$, and divided into annuli of 5$''$  width as was done by T08.  At our convolved resolution, this allows us to Nyquist sample the data.  As an example, Figure~\ref{m51} shows the \ion{H}{1} emission (bottom panel) and H$_{2}$ map (upper panel) in polar coordinates for NGC 5194 with the contours of 24 $\mu$m emission overlaid.  One notes that in both cases the arm patterns in these tracers are nearly coincident.  Thus, any offsets between these tracers must be small.

The cross-correlation coefficients were calculated for angular offsets from -180$^{\circ}$ to +180$^{\circ}$ in increments of $\sim 0.1^{\circ}$.  We restrict the search for the peak between $\pm30^{\circ}$, because the offsets must be at least this small based on visual inspection (see Figure~\ref{m51}).  We fit the region of the maxima with a four degree polynomial and calculate its peak value.  The angular position of the peak is the angular offset, $\Delta \phi (r)$, between the tracers at each radius.  Any local maxima with a cross-correlation coefficient less than 0.3 is rejected, in following T08.  The sign of the angular offsets is corrected for the direction of rotation of the galaxy.  

\subsection{Angular Offsets Between \ion{H}{1} and 24 $\mu$m Images} 

\begin{figure}

\centering

 \includegraphics[trim=25mm 0mm 0mm 250mm, clip,width=80mm]{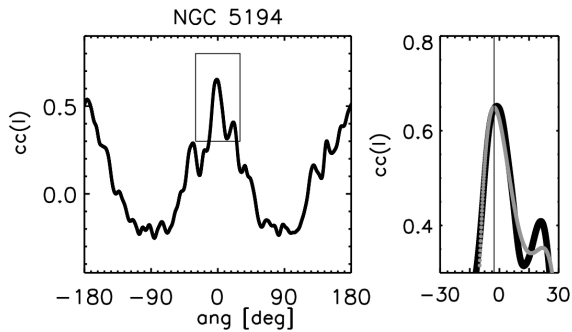}

\caption[Cross-correlation functions $cc(l)$ as a function of azimuth offset for NGC 5194.]{Example of a cross-correlation function, $cc(l)$, for \ion{H}{1} and 24 $\mu$m as a function of azimuthal offset for NGC 5194 at a radius of 80$"$.  Two broad peaks are present due to the self-similarity of the two spiral arms.  The box on the right shows the region zoomed in ($\pm 30^{\circ}$).  The fourth degree polynomial fit overlaid in gray.  The location of the maximum of the polynomial (black vertical line) gives the angular offset between the two tracers.  There is negligible offset between these tracers at this radius (2.7$^{\circ}$).  This example is not atypical for different galaxies, and radii.}

\label{singlecc}

\end{figure}

Figure~\ref{singlecc} shows a sample cross-correlation function between \ion{H}{1} and 24 $\mu$m at a radial annulus of 80$''$ for NGC 5194.  One notes two broad peaks due to the two spiral arms and the self-similarity of the structure.  The zoomed box on the right shows the peak and the polynomial fit (gray curve), the maximum of which locates the angular offset at that radius (vertical black line).  The offsets are measured in a similar fashion at all radii.


\begin{figure}

\centering

 \includegraphics[trim=140mm 0mm 10mm 200mm, clip,width=100mm]{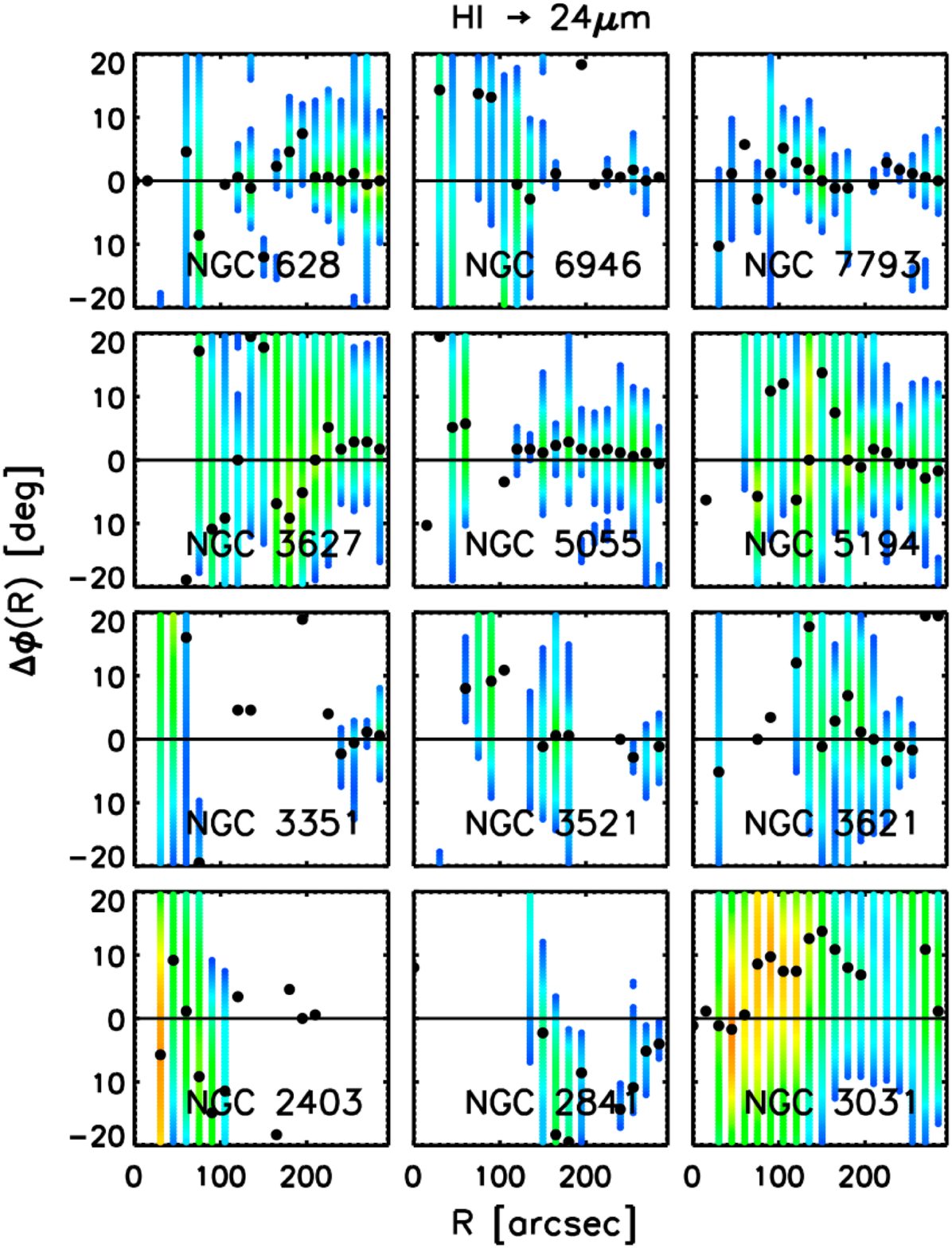}

\caption[Radial profiles of the angular offsets based on the location of the peak of the cross-correlation function between \ion{H}{1} and 24 $\mu$m versus radius.]{Radial profiles of the angular offsets based on the location of the peak of the cross-correlation function between \ion{H}{1} and 24 $\mu$m vs. radius (black dots).  The colors reflect the value of the cross-correlation coefficients from high (red) to low (blue) in the range of $\pm 20^{\circ}$.  Values below 0.3, our chosen cutoff, are not shown.  The angular offsets are corrected for the direction of rotation.  None of the galaxies show the trend predicted by the R69 model.}

\label{allh1}

\end{figure}

Figure~\ref{allh1} shows the angular offsets (black points) versus radius for all 12 galaxies.  The colors reflect the value of the cross-correlation with low values ($\approx0.3$) being blue and high values ($\approx0.8$) being red in the neighborhood of the chosen offset corresponding to the peak value ($\pm 20^{\circ}$).  Around the peak value, there is a broad region where $cc(l)$ is high.  One expects to find positive angular offsets smoothly decreasing with radius and crossing zero at corotation.  
The radial profiles of the angular offsets do not show any smooth trend of going from positive to negative values and show considerable scatter.  The profiles do not correspond in any way to the model predictions.  Our profiles also do not agree with the profiles found by T08, who found offsets that agreed with the picture of R69.   In \S4.3 we examine this in detail.  However, if the reader is not interested in this discussion, the \S4.3 can be skipped.

\subsection{Angular Offsets Between H$_{2}$ and 24 $\mu$m, UV and 3.6 $\mu$m Images}
The \ion{H}{1} distribution is not as concentrated to the spiral arms as H$_{2}$ ({\it e.g.}, Foyle et al.\ 2010).  Since the \ion{H}{1} emission is more evenly distributed than the 24 $\mu$m emission, this may make the measurements of angular offsets more difficult. Thus, we cross-correlate H$_{2}$ maps with other SF tracers to measure angular offsets.

\begin{figure}

\centering

 \includegraphics[trim=140mm 0mm 0mm 300mm, clip,width=100mm]{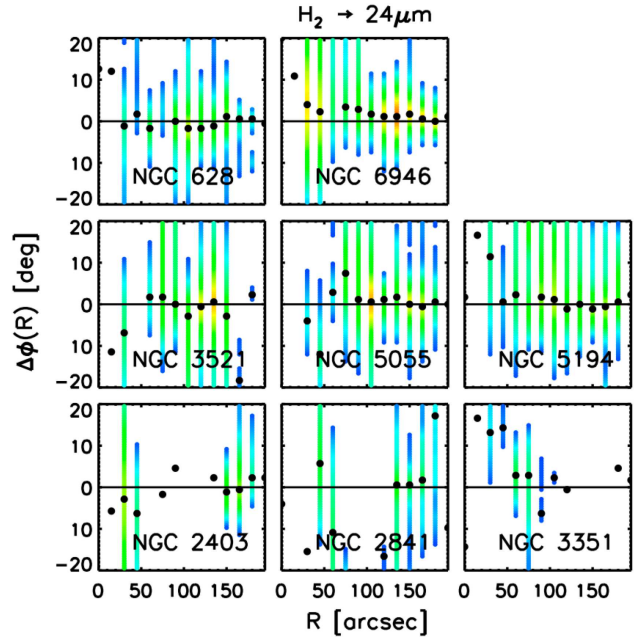}

\caption[Radial profiles of the angular offsets based on the location of the peak of the cross-correlation function between H$_{2}$ and 24 $\mu$m versus radius.]{Same as Figure~\ref{allh1}, but showing the location of the peak of the cross-correlation function between H$_{2}$ and 24 $\mu$m emission. }

\label{allco}

\end{figure}

Eight of the galaxies had CO coverage from HERACLES at 13$''$ resolution.  As described in Foyle et al.\ (2010), we process and convert those maps to produce deprojected maps of H$_{2}$ using an $X_{CO}$ conversion factor.  We then transform the images to polar coordinates.  As we did with \ion{H}{1} and 24 $\mu$m emission, we measure angular offsets between the H$_{2}$ and 24 $\mu$m emission by locating the peak of the cross-correlation function.  Figure~\ref{allco} shows the radial profile of the angular offsets between these two tracers.  In a similar fashion to Figure~\ref{allh1}, we find little evidence for systematic offsets, in contrast to R69 and T08.  There is some evidence for a sequence of offsets in NGC 6946 and NGC 5194.  Thus, we examine these galaxies as well as NGC 628 in greater detail and with other tracers for possible angular offsets.  

For three galaxies with prominent spiral structure we added FUV images from {\it GALEX} (Gil de Paz et al.\ 2007) which trace a longer timescale and should show the greatest offsets. We also included 3.6 $\mu$m emission maps which trace the underlying old stellar population.  We measured angular offsets between the H$_{2}$ and the UV emission as well as H$_{2}$ and the 3.6 $\mu$m emission using the cross-correlation method described above.  The timescale between the H$_{2}$ and the UV is similar to the timescale between the gas and the 100 Myr clusters in the simulations of DP10 (see \S3).   

\begin{figure}

\centering

 \includegraphics[trim=35mm 0mm 10mm 50mm, clip,width=80mm]{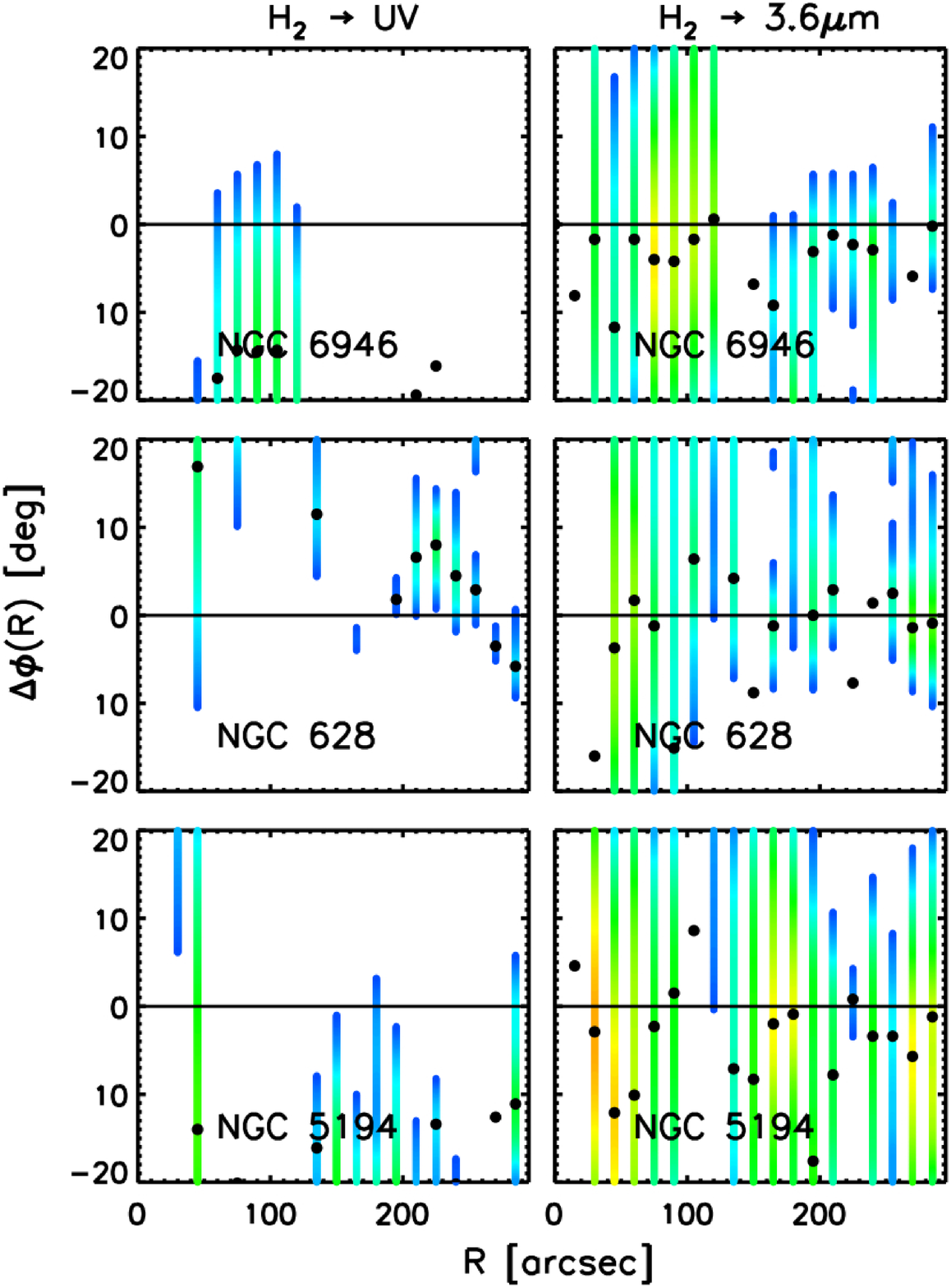}

\caption[Radial profiles of the angular offsets based on the location of the peak of the cross-correlation function between other tracers]{Same as Figure~\ref{allh1}, but showing the location of the peak of the cross-correlation function between H$_{2}$ and UV emission (left) and H$_{2}$ and 3.6 $\mu$m emission (right). }

\label{three}

\end{figure}

Figure~\ref{three} shows the angular offset profiles for cross-correlations of H$_{2}$ with the UV emission on the left and the H$_{2}$ with the 3.6 $\mu$m emission on the right.  Of these galaxies, only NGC 628 presents any trend suggestive of the picture of R69.  However, it is important to note that many of these offsets are well below the resolution of the images, especially for the case of H$_{2}$, which only has 13$''$ resolution.  We also tried other combinations of these images and divided the images into quadrants to isolate individual arms.  Our findings were similar in all cases.

\subsection{Comparison with Tamburro et al.\ (2008)}
\begin{figure}

\centering

 \includegraphics[trim=5mm 0mm 0mm 0mm, clip,width=80mm]{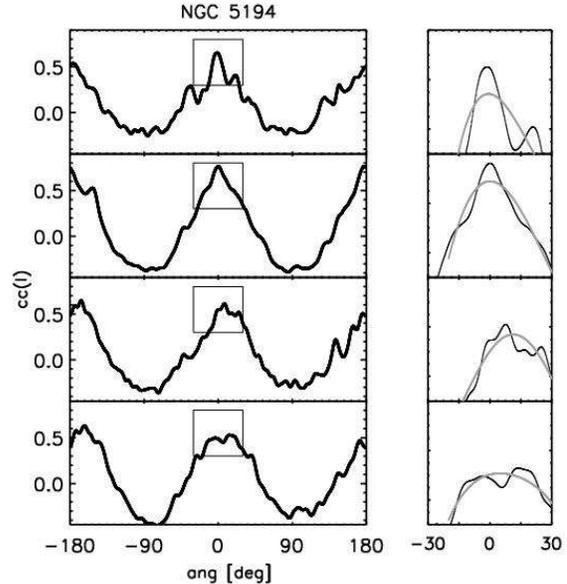}

\caption[Cross-correlation functions $cc(l)$ as a function of azimuth offset for NGC 5194.]{Cross-correlation functions $cc(l)$ between \ion{H}{1} and 24 $\mu$m as a function of azimuth offset for NGC 5194.  Each panel represents neighboring annuli from 60$''$ (bottom) to 80$''$ (top).  The right column shows the boxed region zoomed in ($\pm 30^{\circ}$).  The fourth degree polynomial fits are overlaid in gray.  The peak is often quite broad and bumpy and the polynomial peak depends heavily on the range of points selected.}

\label{cc51}

\end{figure}

We have found no evidence for a systematic ordering of angular offsets in the way found by T08.  We examine the cause of this discrepancy in greater detail here.  Figure~\ref{cc51} shows the cross-correlation functions of \ion{H}{1} with 24 $\mu$m emission at several radial annuli for NGC 5194.  The left panels show the full range from -180$^{\circ}$ to +180$^{\circ}$.  We also zoom in on the peak of the function in the boxed regions and display these on the right.  The zoomed regions reflect the range of -30$^{\circ}$ to +30$^{\circ}$.  The grey curve shows the fourth degree polynomial fit.

 The top panel of  Figure~\ref{cc51} exhibits what one would hope to find when seeking the maximum of $cc(l)$.  At this radius a single, narrow peak is found.  The other panels, show, however, that at most radii, the $cc(l)$ is more complex, with a very broad peak and uneven features.  The polynomial fit is sensitive to the range over which one fits and selecting this range can be subjective.   Due to this, we also tried selecting the maximum $cc(l)$ value, rather than fitting the peak.  Even in this case, no systematic variation of offsets were found.  We tried a number of variations of T08's method, including smoothing and filtering of the images, variations of the cross-correlation calculation and restrictions of narrow regions around the arms in the case of non-axisymmetric distributions of stars and gas and were unable to reproduce the findings of T08.

\begin{figure}

\centering

 \includegraphics[trim=35mm 40mm 30mm 50mm, clip,width=60mm]{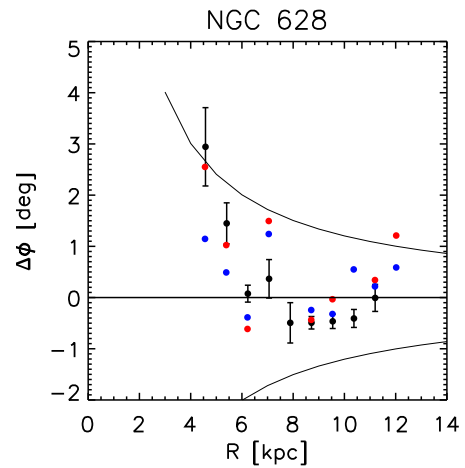}

\caption[Comparison of the offsets found by T08 and the offsets measured in this study.]{Comparison of the offsets found by T08 (black) and the offsets measured in this study using two slightly different fitting ranges of the $cc(l)$ (blue and red).  The offsets measured by T08 are all below the angular resolution limit, which varies with radius (thin black line).}

\label{dt}

\end{figure}

We carried out a careful comparison of NGC 628 also studied by T08.  While we were able to reproduce some of their offsets, the offsets measured were sensitive to small changes in the polynomial fit and the way in which the cross-correlation function is calculated ({\it i.e.} use of mean or median for $\bar{x}$ and $\bar{y}$ in Eq.~\ref{ccl}).  Figure~\ref{dt} shows the offsets found by T08 (black) and two of our attempts using different fitting ranges for the polynomial (blue and red).  While T08's offsets show a clear progression from high values to low values, which cross zero, our points are much more scattered.  In part, we can attribute this to the fact that T08 selected the peak to fit and chose the range over which they fit by-eye.  We also tried to choose specific fitting ranges at each radius, but found the choice to be subjective.  
 
Beyond the sensitivity of the peak position to the range of the polynomial fit, resolution may be a cause for the lack of agreement with T08.  The angular resolution varies with radius and the thin black line in Figure~\ref{dt} shows that all of the offsets found by T08 are below the resolution of the image.  The common image resolution is 6$''$ and at a radius of 30$''$ this corresponds to an angle of 11$^{\circ}$.  Most of the offsets found by T08 were less than 5$^{\circ}$ except in the very inner regions, where resolution concerns are even greater. However, given that the cross-correlation function is calculated considering all azimuthal positions, it is possible that one can probe below the resolution limit to measure the offsets.  The fact that T08's offsets are not randomly scattered points to this possibility.

\begin{figure}

\centering

 \includegraphics[trim=0mm 0mm 0mm 0mm, clip,width=80mm]{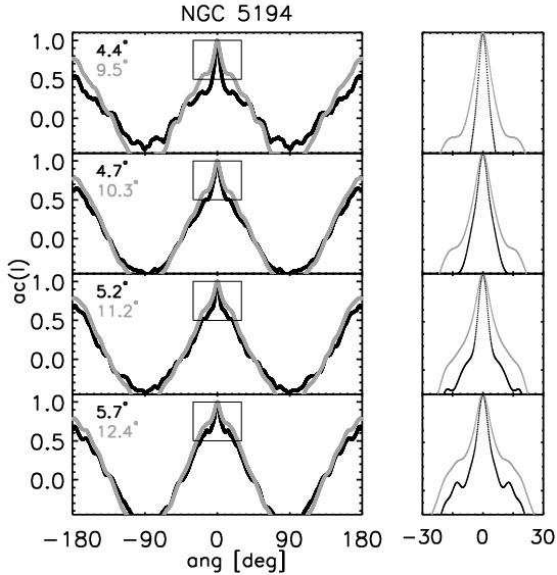}

\caption[The effect of spatial resolution on the autocorrelation function of
\ion{H}{1} maps at two different resolutions.]{Effect of spatial resolution on the
  autocorrelation function of \ion{H}{1} maps with 6$''$ (black) and 13$''$ (gray)
  resolution for NGC 5194 at a series of fiducial radii.  Since the azimuthal
  resolution varies with radius, it is listed in the
  upper left of each panel.  As one moves out in radius (from bottom to top)
  the azimuthal resolution size increases.  As in Figure~\ref{cc51} we zoom in on the
  peak between $\pm$ 30$^{\circ}$.  We see that as the size of azimuthal
  resolution increases, the peak of $cc(l)$ broadens and increases in value.}

\label{h1res}

\end{figure}

It is not clear how resolution might affect the cross-correlation function and
the determination of the peak.  We first examine how resolution affects the
autocorrelation function of two \ion{H}{1} maps at a resolution of 6$''$ and 13$''$
using NGC 5194.  The autocorrelation function is calculated like the $cc(l)$
(Eq.~\ref{ccl}), but in this case both $x$ and $y$ represent the same \ion{H}{1} map.  For
both resolutions, we expect the peak of the autocorrelation function to be
centered at zero with a maximum of unity.  Figure~\ref{h1res} shows the autocorrelation function for a
series of annuli.  Since the azimuthal resolution varies with radius, we list
the azimuthal resolution for each pair of images in the upper left of the
panels.  We see that, as expected, the autocorrelation functions have a peak at zero and maxima of unity.  However, one notes that as the resolution size increases, the autocorrelation
function broadens and has a higher coefficient over a greater range of angles.  We divided each annulus into
units smaller than the resolution. Thus, as the resolution size increases, the
fraction of pixels with identical values increases.  This provides a better
agreement in the cross-correlation, which broadens and increases the peak.

\begin{figure}

\centering

 \includegraphics[trim=0mm 0mm 0mm 0mm, clip,width=80mm]{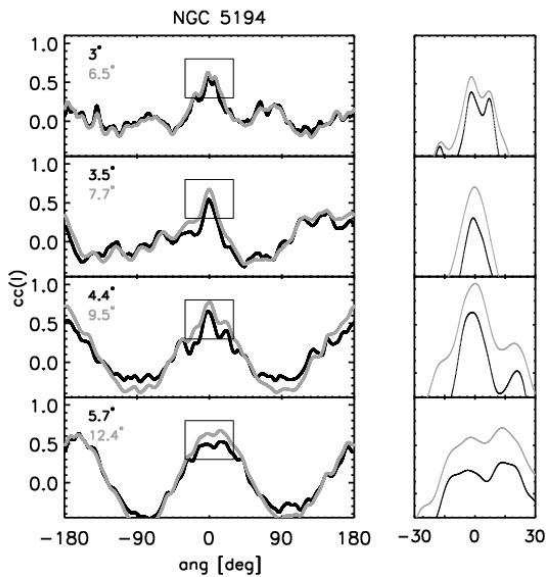}

\caption[The effect of spatial resolution on the $cc(l)$ function using \ion{H}{1} and
24 $\mu$m images.]{Effect of spatial resolution on the $cc(l)$ function
  calculated using \ion{H}{1} and 24 $\mu$m maps with 6$''$ (black) and 13$''$ (gray)
  resolution for NGC 5194 at a series of fiducial radii.  Since the azimuthal
  resolution varies with radius, it is listed in the upper left of each
  panel.  As one moves out in radius (from bottom to top) the azimuthal
  resolution size increases.  As in Figure~\ref{cc51} we zoom in on the peak between $\pm$ 30$^{\circ}$.  While the $cc(l)$ is roughly similar, features on the peak change substantially, which may introduce differences in the polynomial fit and the position of the maxima.}

\label{compres}

\end{figure}

In order to examine how the peak position might vary with resolution, we consider the cross-correlation of \ion{H}{1} and 24 $\mu$m at two different resolutions in Figure~\ref{compres}.  We use pairs of images at our common resolution of 6$''$ (black), and 13$''$ (gray).  Like the autocorrelation function, as the resolution size increases, the $cc(l)$ broadens and has a higher value.  The functions are very similar in shape, but as one can see in the zoomed in regions, when the resolution size is large (small radii), the shape of the peak can change substantially.  This affects the polynomial fit and ultimately the position of the peak center.  In the inner regions, where the resolution effects are greatest, we found that the peak positions varied by as much as 8$^{\circ}$.  However, at larger radii the difference was less than a degree and the median deviation was 0.5$^{\circ}$.  Thus, while the peak position remains roughly the same, it is important to consider the uncertainties introduced by resolution in determining its precise position.

In summary, we can attribute our lack of agreement with T08 to a combination of three causes: 1) the offset value is very sensitive to the polynomial fitting of the peak; 2) the offset value is very sensitive to small changes in the way the $cc(l)$ function is calculated; and 3) almost all of the offsets are below the resolution limit of the images and we have found that changes in the resolution of the image do affect the shape of the $cc(l)$ function particularly at small radii.


\section{Comparison of Simulations and Observations}
For all galaxies in the observed sample and between all tracers we found a clear peak in the cross-correlation function (see Fig~\ref{singlecc}).  The location of the peak marks the angular offset between the two tracers considered.  However, the measured angular offsets showed no systematic variation as would be expected from a long-lived, quasi-stationary spiral structure.  Among the simulated galaxies only the one with an imposed stationary potential showed clear offsets with a systematic variation.  The barred, interacting and flocculent transient spirals all showed very small offsets (less than a few degrees) and the interacting and flocculent spirals showed no clear systematic variation.  This agrees well with what we found in the observations for the various tracers of gas and recent SF.  The absence of observed offsets not only suggests that these galaxies do not have a static spiral pattern, but also that their structure is  unlikely to be the result of a single mode transient spiral that persists for longer times.

However, upon comparing the shape of the cross-correlation functions of the simulated galaxies, with those observed, we do not find a simulated model that agrees in detail.  The observations tend to show at least two broad peaks, depending on the nature and number of arms in the spiral pattern (see Fig~\ref{singlecc}) .  The simulated barred galaxy, did show two clear peaks, but the cross-correlation coefficients were much smaller than those seen in observations.   In the case of the flocculent and interacting spiral simulations, we did not find such clear features.  It is interesting to compare the cross-correlation function shown in Fig~\ref{singlecc} with that of the interacting, simulated spiral galaxy in Fig~\ref{DBcc}.  The interacting spiral is structurally similar to NGC 5194.  While the cross-correlation function of NGC 5194 showed two broad peaks, the simulated galaxy showed only one.   One possible cause of the discrepancy in the shape cross-correlation function in the simulations and observations is because we only have a relatively small number of discrete clusters in the simulations, rather than the more continuous map of emission in the observation.

\section{Conclusions}
We have used simulations from DP10 and high resolution observations for a sample of 12 spiral galaxies to look for evidence for a model of spiral structure that is long-lived, quasi-stationary with a constant pattern speed.  This model predicts that SF tracers should be offset from one another reflecting the relative velocity between the disk and pattern and the onset time for SF.

We have used an algorithmic technique developed by T08 to measure angular offsets between the gas and stellar clusters in four simulated galaxies with different spiral structures including a galaxy with a stationary spiral potential, a barred galaxy, an interacting galaxy and a more flocculent, transient spiral structure.  Only when a stationary spiral potential was imposed did we find clear angular offsets ordered in the way predicted by a long-lived, stable structure.  For the barred galaxy, though we found angular offsets for younger clusters, and some indication of an age transition, the offsets were very small ($<$5$^{\circ}$), and thus inconclusive. For the interacting and transient spirals, there were no clear cross correlation peaks and no measurable offsets.   

We then used the same method to measure angular offsets between the gas and SF tracers in a sample of 12 observed galaxies.  In all cases we did not have a systematic ordering in the way predicted by the simple prescription of R69.  We specifically measured offsets between \ion{H}{1} and 24 $\mu$m emission in order to directly compare with the work of T08 and are unable to reproduce their results (see \S4.3 for a discussion).  This study has also measured offsets between CO, which traces the molecular gas with the 24 $\mu$m emission.  Even in this case, which directly probes the time between molecular cloud formation and SF, we find no systematic ordering of angular offsets.  For three galaxies we also examined the cross-correlation of other tracer combinations including UV and 3.6 $\mu$m and found similar results.  

Our results contradict previous by-eye estimates of such offsets.  While there may be some patches of SF
tracers that show by-eye offsets, an algorithmic analysis shows that such patches are isolated and there is no overall trend which supports the prediction of a model where the spiral structure is long-lived with a constant pattern speed.  Our technique made use of the highest quality images to date and, provided the timescales of SF are at least a few Myr, angular offsets should have been detectable.  There are number of possible reasons why such a systematic ordering was not detected.  It may be the case that there are no angular offsets between the tracers.  This would be the case if the spiral structure were simply a result of sheared patches of star forming regions or if the structure was rapidly dissolving and reforming.  However, it is also possible that offsets are present, but that structure is more complex than a single pattern with a constant pattern speed.  Indeed, recent studies have measured multiple pattern speeds for several spiral galaxies(Meidt et al.\ 2009).  Furthermore,  there is substantial emission from SF tracers in the interarm regions (at least 30\%) (Foyle et al.\ 2010).  This may be obfuscating the offsets near the spiral arms due to our cross-correlation method which uses all azimuthal positions.     The arms may also have slightly different offsets and the stars may have highly elliptical orbits, both of which would introduce uncertainties when fitting over the whole galaxy as our method relies on a polar coordinate system. Thus, there is still room for the possibility that  systematic offsets between SF tracers exists in the arms, but we may not be able to quantitatively measure them with this technique.   Furthermore, the fact a cross-correlation peak could be found at most radii, suggests that angular offset measurements could still be used to understand star-forming sequences if the local dynamics are known.

The fact that there is little evidence for angular offsets between SF tracers as predicted by the model of a long-lived spiral structure with a constant pattern speed lends support to a model with a transient spiral structure that reorganizes the interstellar medium ({\it e.g.}, Elmegreen \& Elmegreen 1986; Dobbs \& Bonnell 2008; Sellwood 2010) or models that have spiral arms forming due to the shearing of gas and stars by differential rotation ({\it e.g.},  Seiden \& Gerola 1979 and Elmegreen et al.\ 2003).  Simulations largely support such pictures and other observational studies have shown that most spiral structures are quite complex with multiple pattern speeds ({\it e.g.}, Meidt et al.\ 2009).  We caution, however, that the cross-correlation functions of the other simulated spiral structures in this study, did not agree in detail with the observations.  Thus, continued detailed comparisons between observations and simulations will be required in order to uncover the nature and persistence of spiral structure.

\acknowledgements{K.Foyle acknowledges generous support from the Max Planck Society, International Max Planck Research School for Astronomy and Cosmic Physics at the University of Heidelberg. We thank both Karl Schuster and the referee, for their comments, which greatly improved this work.}


\clearpage

\begin{thebibliography}{}



\bibitem[Bertin et al. (1989)]{1989ApJ...338...78B} Bertin, G., Lin, C.~C., Lowe, S.~A., \& Thurstans, R.~P. 1989, \apj, 338, 78 
\bibitem[Bertin et al. (1989)]{1989ApJ...338..104B} Bertin, G., Lin, C.~C., Lowe, S.~A., \& Thurstans, R.~P. 1989, \apj, 338, 104 
\bibitem[Binney \& Tremaine(2008)]{2008gady.book.....B} Binney, J., \& Tremaine, S.  2008, Galactic Dynamics: Second Edition, Princeton University Press, Princeton, NJ USA, 2008.
\bibitem[Calzetti et al.(2007)]{2007ApJ...666..870C} Calzetti, D., et al.\ 2007, \apj, 666, 870

\bibitem[Dobbs et al. (2006)]{2006MNRAS.371.1663D} Dobbs, C.~L., Bonnell, I.~A., \& Pringle, J.~E. 2006, \mnras, 371, 1663
\bibitem[Dobbs(2007)]{2007PhDT.........1D} Dobbs, C. 2007, Ph.D.~Thesis, University of St Andrews Physics \& Astronomy theses
 
\bibitem[Dobbs \& Bonnell(2008)]{2008MNRAS.385.1893D} Dobbs, C.~L., \& Bonnell, I.~A. 2008, \mnras, 385, 1893
 
\bibitem[Dobbs \& Pringle(2010)]{2010arXiv1007.1399D} Dobbs, C.~L., \& Pringle, J.~E. 2010, arXiv:1007.1399

\bibitem[Egusa et al. (2009)]{2009ApJ...697.1870E} Egusa, F., Kohno, K., Sofue, Y., Nakanishi, H., \& Komugi, S. 2009, \apj, 697, 1870  
\bibitem[Elmegreen 
\& Elmegreen(1986)]{1986ApJ...311..554E} Elmegreen, B.~G., \& Elmegreen, D.~M.\ 1986, \apj, 311, 554 
\bibitem[Elmegreen et al.(2003)]{2003ApJ...590..271E} Elmegreen, B.~G., Elmegreen, D.~M., \& Leitner, S.~N. 2003, \apj, 590, 271 
\bibitem[Foyle et al. (2010)]{2010ApJ...725..534F} Foyle, K., Rix, H.-W., Walter, F., \& Leroy, A.~K. 2010, \apj, 725, 534 
\bibitem[Fujii et al. (2010)]{2010arXiv1006.1228F} Fujii, M.~S., Baba, J., Saitoh, T.~R., Makino, J., Kokubo, E., \& Wada, K. 2010, arXiv:1006.1228 

\bibitem[Garcia-Burillo et al. (1993)]{1993A&A...274..123G} Garcia-Burillo, S., Guelin, M., \& Cernicharo, J. 1993, \aap, 274, 123 
\bibitem[Gil de Paz et al. (2007)]{2007ApJS..173..185G} Gil de Paz, A., et al.\ 2007, \apjs, 173, 185
\bibitem[Gittins \& Clarke(2004)]{2004MNRAS.349..909G} Gittins, D.~M., \& Clarke, C.~J. 2004, \mnras, 349, 909 
\bibitem[Goldreich \& Lynden-Bell(1965)]{1965MNRAS.130..125G} Goldreich, P., \& Lynden-Bell, D. 1965, \mnras, 130, 125
\bibitem[Helfer et al. (2003)]{2003ApJS..145..259H} Helfer, T.~T., Thornley, M.~D., Regan, M.~W., Wong, T., Sheth, K., Vogel, S.~N., Blitz, L., \& Bock, D.~C.-J.  2003, \apjs, 145, 259
\bibitem[Humphreys \& Sandage(1980)]{1980ApJS...44..319H} Humphreys, R.~M., \& Sandage, A. 1980, \apjs, 44, 319
\bibitem[Julian \& Toomre(1966)]{1966ApJ...146..810J} Julian, W.~H., \& Toomre, A. 1966, \apj, 146, 810
\bibitem[Kennicutt et al. (2003)]{2003PASP..115..928K} Kennicutt, R.~C., Jr., et al.\ 2003, \pasp, 115, 928
 
\bibitem[Kranz et al.(2003)]{2003ApJ...586..143K} Kranz, T., Slyz, A., \& Rix, H.-W. 2003, \apj, 586, 143 
\bibitem[Leroy et al.(2008)]{2008AJ....136.2782L} Leroy, A.~K., Walter, F., Brinks, E., Bigiel, F., de Blok, W.~J.~G., Madore, B., \& Thornley, M.~D.\ 2008, \aj, 136, 2782 
\bibitem[Leroy et al.(2009)]{2009AJ....137.4670L} Leroy, A.~K., et al.\ 2009, \aj, 137, 4670 

\bibitem[Lin \& Shu(1964)]{1964ApJ...140..646L} Lin, C.~C., \& Shu, F.~H. 1964, \apj, 140, 646
\bibitem[Lin \& Shu(1966)]{1966PNAS...55..229L} Lin, C.~C., \& Shu, F.~H. 1966, Proceedings of the National Academy of Science, 55, 229
\bibitem[Loinard et al.(1996)]{1996ApJ...469L.101L} Loinard, L., Dame, T.~M., Koper, E., Lequeux, J., Thaddeus, P., \& Young, J.~S. 1996, \apjl, 469, L101 
\bibitem[Lord \& Young(1990)]{1990ApJ...356..135L} Lord, S.~D., \& Young, J.~S. 1990, \apj, 356, 135
\bibitem[Lubow et al.(1986)]{1986ApJ...309..496L} Lubow, S.~H., Cowie, L.~L., \& Balbus, S.~A. 1986, \apj, 309, 496
\bibitem[Lynden-Bell \& Kalnajs(1972)]{1972MNRAS.157....1L} Lynden-Bell, D., \& Kalnajs, A.~J. 1972, \mnras, 157, 1 
\bibitem[Lynds(1970)]{1970IAUS...38...26L} Lynds, B.~T. 1970, The Spiral Structure of our Galaxy, 38, 26
\bibitem[Mathewson et al.(1972)]{1972A&A....17..468M} Mathewson, D.~S., van der Kruit, P.~C., \& Brouw, W.~N. 1972, \aap, 17, 468 
\bibitem[Meidt et al.(2009)]{2009ApJ...702..277M} Meidt, S.~E., Rand, R.~J., \& Merrifield, M.~R. 2009, \apj, 702, 277

\bibitem[Peng et al.(2002)]{2002AJ....124..266P} Peng, C.~Y., Ho, L.~C., Impey, C.~D., \& Rix, H.-W. 2002, \aj, 124, 266 
\bibitem[Rand \& Kulkarni(1990)]{1990ApJ...349L..43R} Rand, R.~J., \& Kulkarni, S.~R. 1990, \apjl, 349, L43 
\bibitem[Rand(1995)]{1995AJ....109.2444R} Rand, R.~J. 1995, \aj, 109, 2444 
\bibitem[Roberts(1969)]{1969ApJ...158..123R} Roberts, W.~W. 1969, \apj, 158, 123 
\bibitem[Roberts et al.(1975)]{1975ApJ...196..381R} Roberts, W.~W., Jr.,  Roberts, M.~S., \& Shu, F.~H. 1975, \apj, 196, 381
\bibitem[Rots(1975)]{1975A&A....45...43R} Rots, A.~H. 1975, \aap, 45, 43 
\bibitem[Seiden \& Gerola(1979)]{1979ApJ...233...56S} Seiden, P.~E., \& Gerola, H. 1979, \apj, 233, 56 
\bibitem[Sellwood \& Carlberg(1984)]{1984ApJ...282...61S} Sellwood, J.~A., \& Carlberg, R.~G. 1984, \apj, 282, 61
\bibitem[Sellwood(2010)]{2010arXiv1001.5430S} Sellwood, J.~A. 2010, arXiv:1001.5430 
\bibitem[Shetty et al.(2007)]{2007ApJ...665.1138S} Shetty, R., Vogel, S.~N., Ostriker, E.~C., \& Teuben, P.~J.\ 2007, \apj, 665, 1138
\bibitem[Tamburro et al.(2008)]{2008AJ....136.2872T} Tamburro, D., Rix, H.-W., Walter, F., Brinks, E., de Blok, W.~J.~G., Kennicutt, R.~C., \& MacLow, M.-M. 2008, \aj, 136, 2872 
\bibitem[Thomasson et al.(1990)]{1990ApJ...356L...9T} Thomasson, M., Elmegreen, B.~G., Donner, K.~J., \& Sundelius, B. 1990, \apjl, 356, L9 
\bibitem[Toomre(1981)]{1981seng.proc..111T} Toomre, A. 1981, Structure and Evolution of Normal Galaxies, 111 
\bibitem[Vogel et al.(1988)]{1988Natur.334..402V} Vogel, S.~N., Kulkarni, S.~R., \& Scoville, N.~Z. 1988, \nat, 334, 402
\bibitem[Wada(2008)]{2008ApJ...675..188W} Wada, K. 2008, \apj, 675, 188
\bibitem[Walter et al.(2008)]{2008AJ....136.2563W} Walter, F., Brinks, E., de Blok, W.~J.~G., Bigiel, F., Kennicutt, R.~C., Thornley, M.~D., \& Leroy, A. 2008, \aj, 136, 2563



\end{thebibliography}
\end{document}